\documentclass[aip,
 mph,%
 amsmath,
 amssymb,
 superscriptaddress,
 floatfix,
 preprint,
 eqsecnum,%
 numerical,%
]{revtex4-1}

\usepackage{graphicx}
\DeclareGraphicsExtensions{.pdf,.png,.jpg}
\usepackage[caption=false%,labelformat=parens,subrefformat=subparens,listofformat=subparens
]{subfig}
\usepackage{epstopdf}
\usepackage{lipsum}
\usepackage{wrapfig}
\usepackage{wasysym}
\usepackage{threeparttable,booktabs,siunitx}
\usepackage{booktabs}
\usepackage{lmodern}
\usepackage{longtable}
\usepackage{multirow}
\usepackage{amsmath}
\usepackage[OT2,T1]{fontenc}
\usepackage[bookmarks=false]{hyperref}
\usepackage{color}
\hypersetup{pdfauthor={},pdfborder=0 0 0,colorlinks=true,citecolor=blue,linkcolor=blue,urlcolor=blue,}

\DeclareSymbolFont{cyrletters}{OT2}{wncyr}{m}{n}
\DeclareMathSymbol{\comb}{\mathalpha}{cyrletters}{"58}

\newcommand{\ben}{\begin{eqnarray}\displaystyle}
\newcommand{\een}{\end{eqnarray}}

\begin{document}

\title{X-ray differential phase contrast imaging on asymmetric dual-phase grating interferometer with source grating: theory and experiment}

\author{Yongshuai Ge}%
\thanks{Yongshuai Ge and Jianwei Chen have made equal contributions to this work and both are considered as the first authors.}
\affiliation{Paul C Lauterbur Research Center for Biomedical Imaging, Shenzhen Institutes of Advanced Technology, Chinese Academy of Sciences, Shenzhen, Guangdong 518055, People's Republic of China}
\affiliation{Research Center for Medical Artificial Intelligence, Shenzhen Institutes of Advanced Technology, Chinese Academy of Sciences, Shenzhen, Guangdong 518055, People's Republic of China}
\affiliation{Chinese Academy of Sciences Key Laboratory of Health Informatics, Shenzhen, Guangdong 518055, People's Republic of China}
\author{Jianwei Chen}%
 \thanks{Yongshuai Ge and Jianwei Chen have made equal contributions to this work and both are considered as the first authors.}
\affiliation{Paul C Lauterbur Research Center for Biomedical Imaging, Shenzhen Institutes of Advanced Technology, Chinese Academy of Sciences, Shenzhen, Guangdong 518055, People's Republic of China}
\affiliation{Research Center for Medical Artificial Intelligence, Shenzhen Institutes of Advanced Technology, Chinese Academy of Sciences, Shenzhen, Guangdong 518055, People's Republic of China}
\author{Peiping Zhu}
\affiliation{Platform Advanced Photon Source Technology R$\&$D (PAPS), Institute of High Energy Physics, Chinese Academy of Sciences, Beijing, 100049, People's Republic of China}
\affiliation{University of Chinese Academy of Sciences, Beijing, 100049, People's Republic of China}
\author{Jun Yang}
\affiliation{Key Laboratory of Optoelectronic Devices and Systems of Ministry of Education and Guangdong Province, College of physics and Optoelectronic Engineering, Shenzhen University, Shenzhen 518060, People's Republic of China}
\author{Ronghui Luo}
\affiliation{Research Center for Medical Artificial Intelligence, Shenzhen Institutes of Advanced Technology, Chinese Academy of Sciences, Shenzhen, Guangdong 518055, People's Republic of China}
\author{Wei Shi}
\affiliation{Paul C Lauterbur Research Center for Biomedical Imaging, Shenzhen Institutes of Advanced Technology, Chinese Academy of Sciences, Shenzhen, Guangdong 518055, People's Republic of China}
\author{Kai Zhang}
\affiliation{Platform Advanced Photon Source Technology R$\&$D (PAPS), Institute of High Energy Physics, Chinese Academy of Sciences, Beijing, 100049, People's Republic of China}
\affiliation{University of Chinese Academy of Sciences, Beijing, 100049, People's Republic of China}
\author{Jinchuan Guo}
\affiliation{Key Laboratory of Optoelectronic Devices and Systems of Ministry of Education and Guangdong Province, College of physics and Optoelectronic Engineering, Shenzhen University, Shenzhen 518060, People's Republic of China}
\author{Hairong Zheng}
\affiliation{Paul C Lauterbur Research Center for Biomedical Imaging, Shenzhen Institutes of Advanced Technology, Chinese Academy of Sciences, Shenzhen, Guangdong 518055, People's Republic of China}
\affiliation{Chinese Academy of Sciences Key Laboratory of Health Informatics, Shenzhen, Guangdong 518055, People's Republic of China}
\author{Dong Liang}
\thanks{Scientific correspondence should be addressed to Dong Liang (dong.liang@siat.ac.cn).}
\affiliation{Paul C Lauterbur Research Center for Biomedical Imaging, Shenzhen Institutes of Advanced Technology, Chinese Academy of Sciences, Shenzhen, Guangdong 518055, People's Republic of China}
\affiliation{Research Center for Medical Artificial Intelligence, Shenzhen Institutes of Advanced Technology, Chinese Academy of Sciences, Shenzhen, Guangdong 518055, People's Republic of China}
\affiliation{Chinese Academy of Sciences Key Laboratory of Health Informatics, Shenzhen, Guangdong 518055, People's Republic of China}

\begin{abstract}
Recently, the dual-phase grating based X-ray differential phase contrast imaging technique has shown better radiation dose efficiency performance than the Talbot-Lau system. In this paper, we provide a theoretical analyses framework derived from wave optics to ease the design of such interferometer systems, including the inter-grating distances, the diffraction fringe period, the phase grating periods, and especially the source grating period if a medical grade X-ray tube with large focal spot is utilized. In addition, a geometrical explanation of the dual-phase grating system similar to the standard thin lens imaging theory is derived with an optical symmetry assumption for the first time. Finally, both numerical and experimental studies have been performed to validate the theory.
\end{abstract}

\maketitle

%%%%%%%%%%%%%%%%%%%%%%%%%%  body  %%%%%%%%%%%%%%%%%%%%%%%%%%
\section{Introduction}
Over the past two decades, the grating-based x-ray interferometry imaging, especially the Talbot-Lau imaging method, has received a huge amount of research interests. In principle, three different images, i.e., the absorption image, the differential phase contrast (DPC) image, and the dark-field (DF) image, with unique contrast mechanism can be generated from the acquired same dataset. Among the above three images, the latter two novel signals are usually considered as complimentary contrast information to the conventional absorption signal. Studies have shown that these two complimentary contrast information may have advancements. For example, the DPC signal, which corresponds to the x-ray refraction information, may have advancements in providing superior contrast sensitivity for certain types of soft tissues\cite{Momose1996, Momose2006, Pfeiffer2006a, Bech2009, Jensen2011a, li2014grating}. Additionally, the DF signal, which corresponds to the small-angle-scattering (SAS) information, is particularly sensitive to certain fine structures such as microcalcifications inside breast tissue\cite{Anton2013, Michel2013, wang2014non, Grandl2015, scherer2016correspondence}.

Aiming at translating such a promising imaging technique into real medical applications, a lot of efforts have been made to overcome the two major potential challenges encountered by the current Talbot-Lau imaging method. The first challenge is the prolonged data acquisition period because of the time consuming phase stepping procedure, and the second challenge is the reduced radiation dose efficiency due to the use of the post-object analyzer grating. During the past decades, different techniques\cite{Zhu2010, miao2013motionless, zanette2012trimodal,Ge2014, Koehler2015,marschner2016helical} have been developed to shorten its data acquisition period. For instance, Zhu et. al. suggested to extract the phase information with only two phase steps. Marschner et. al. developed a fast DPC-CT imaging by integrating the helical scan with the phase stepping procedure. Miao et. al. developed a novel DPC imaging method via a motionless phase stepping procedure using an electrically controlled focal spot moving technique. Ge et. al. proposed a new analyzer grating design which integrates multiple phase stepping procedure together, by doing so, one is able to achieve single-shot DPC imaging. Overall, these innovations were able to either reduce the number of phase steps significantly, or totally eliminating the mechanical phase stepping procedure to realize single-shot DPC imaging. Therefore, the total data acquisition time can be greatly saved down to the similar level as for the conventional absorption imaging. 

To increase the system radiation dose efficiency, recently, experimental studies\cite{miao2015enhancing, Miao2016a, kagias2017dual} have tried to replace the analyzer grating by one or two phase gratings. Because the phase gratings absorb less X-ray photons than the analyzer gratings, therefore, the system radiation dose efficiency could be increased by at least two times. For example, Miao et. al. have experimentally demonstrated the advancements of three nanometer phase gratings based far-field X-ray interferometer in both increasing the phase sensitivity and the radiation dose efficiency. Kagias et. al. also have demonstrated the feasibility of using two phase gratings to realize DPC imaging. Despite of such an advantage, however, the requirement of micro-focus or mini-focus x-ray sources in the above two pioneering experimental works still hurdles the new techniques' practical application, especially in the medical imaging field. To meet the high tube output flux demand, one really needs to consider a medical grade X-ray tube with relative large focal spot size. However, the use of large focal spot X-ray tube degrades the beam coherence, and could significantly decrease the interferometer performance. One possible solution is to add a source grating in front of the large focal spot source. By doing so, both of the low beam flux problem and the low beam coherence problems can be easily solved. This is known as the Lau effect\cite{gori1979lau}, and has already been widely used in the current Talbot-Lau interferometer system. Although there had theoretical explanations to the dual phase grating system\cite{Miao2016a, yan2018quantitative}, to our best knowledge, unfortunately, there has no rigorous theoretical analyses to predict the period of such source grating used for two phase grating interferometer system. Thus, we provide a theoretical analyses framework derived from wave optics to ease the design of a dual-phase grating based interferometer system, for instance, the inter-grating distances, the diffraction fringe period, the phase grating periods, and the source grating period if a large focal spot sized X-ray tube is utilized.

The remains of this paper is organized as following: the section II presents the theoretical analyses foundation, the numerical method, and the experimental method. In the section III, the general representation of the diffraction fringe, the analyzer grating period, fringe visibility, and fringe period are derived. Both numerical validation results and experimental results are presented to verify the theoretical results. We discuss this work in section IV, and finally make a brief conclusion in section V.

\section{Materials and Methods}
\subsection{Theoretical analyses methods}
\begin{figure*}[t]
\centering
\includegraphics[width=0.85\textwidth]{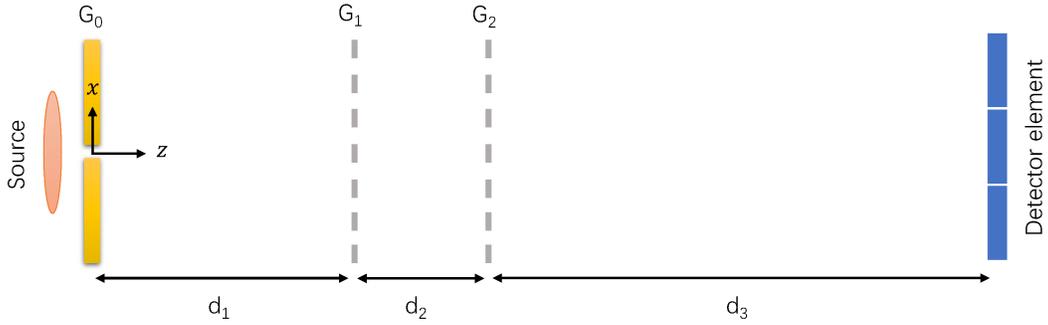}
\vspace{-0.05in}
\caption{Illustration of the assumed imaging geometry. The x-axis is along the vertical direction, and z-axis is along the horizontal direction.}
\label{fig:fig1} 
\end{figure*}

In the theoretical discussion, the assumed imaging geometry is illustrated in Fig.~\ref{fig:fig1}. The distance between the source and the first phase grating is $d_1$, the inter-space between the first and second phase grating is $d_2$, and the distance between the second grating and the detector (observation plane) is $d_3$. The source is assumed as a single slit source:
\begin{align}
 & 	\text{S}(x)=	\text{rect}\left(\frac{x}{\sigma}\right),
\label{eq:source_1}
\end{align}
where $\sigma$ represents the width of the slit opening. This model is chosen because this study mainly focuses on discussing a dual-phase grating system integrated with a source grating. If a micro-focus X-ray tube is used, Eq.~(\ref{eq:source_1}) can be adapted into a Gaussian shaped function.

The diffraction grating is assumed as a periodic transmission function, denoted as $\text{T}(x)$. Its Fourier expansion is modeled as:
\begin{align}
 & 	\text{T}(x)=\sum_{n=-\infty}^{n=\infty}a_{n}e^{\frac{i2\pi nx}{p}},
\label{eq:ch01_trans_f_1}
\end{align}
where $n$ represents the diffraction order, and $p$ denotes the grating period. Depending on the diffraction order of $n$, the coefficient $a_n$ will have different values. By default, the phase gratings are with duty cycle of 0.5. For simplicity, this work only considers two 1D gratings. Nevertheless, this theoretical analysis framework is also capable of dealing with 2D gratings.

In this work, the standard Kirchhoff's diffraction theory\cite{Born1993} is used to facilitate the theoretical analyses of the wave propagation procedure, namely,
\begin{align}
 & 	\text{U}(x)=\frac{\text{U}_{0}e^{ikz}}{i\lambda z}\int\text{S}(\xi)e^{ik\left[\frac{(x-\xi)^{2}}{2z}\right]}d\xi,
\label{eq:ch01_Fresnel_Kirchhoff_4}
\end{align}
where $\text{U}(x)$ denotes the expected X-ray field disturbance at location $(x)$, $\text{U}_{0}$ is the amplitude of the initial disturbance, $\text{S}(\xi)$ represents the generalized X-ray source defined in Eq.~(\ref{eq:source_1}), $\lambda$ denotes the X-ray wavelength, and $z$ is along the wave propagation direction. Essentially, the wave propagation described by Eq.~\ref{eq:ch01_Fresnel_Kirchhoff_4} corresponds to a 1D convolution procedure.

\subsection{Numerical simulation methods}
To quantitatively demonstrate the theoretical results, numerical simulations are performed in Matlab (The MathWorks Inc., Natick, MA, USA) to investigate the fringe period and the visibility. Polychromatic X-ray beam with certain spectrum was considered. The X-ray spectrum was obtained from SpekCalc\cite{poludniowski2009spekcalc} at 40.00 kV with 1.00 mm Al filtration. Two silicon based phase gratings, one is 4.364 $\mu m$ and the other is 4.640 $\mu m$, with 0.5 duty cycle were simulated. The two gratings are designed as $\pi$-phase gratings for X-ray beams with 28 keV energy. For X-ray photons with different energies, the resultant phase shift were calculated correspondingly. The detector pixel has a dimension of 14.00 $\mu m$. In total, 20 independent slit sources with opening width of $8.4 \mu m$ and neighboring distance of $24 \mu m$ are simulated. The source to $G_1$ grating distance is 527.86 mm, and the $G_1$ to $G_2$ distance is 108.90 mm. Diffraction signals are recorded with $d_3$ length varied from 1400.00 mm to 1800.00 mm with step interval of 10.00 mm.

\subsection{Experiments}
\begin{figure*}[t]
\centering
\includegraphics[width=0.85\textwidth]{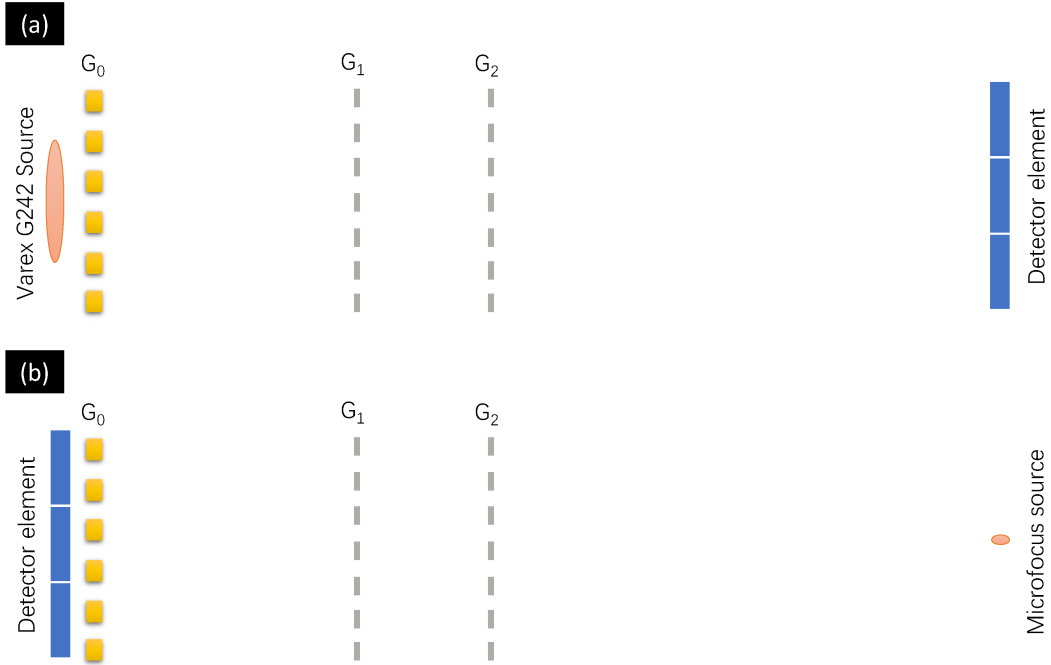}
\vspace{-0.05in}
\caption{Illustration of the two symmetrical experimental setups using dual-phase grating: (a) with the Varex G242 source of $0.4 mm$ nominal focal spot, (b) with the Oxford 96000 microfocus source $0.08 mm$ focal spot.}
\label{fig:Fig_DualPhase_1} 
\end{figure*}
Experimental validations were performed on an in-house X-ray interferometry bench. The system includes a rotating-anode Tungsten target medical grade diagnostic X-ray tube (Varex G242, Varex Imaging Corporation, UT, USA), see Fig.~\ref{fig:Fig_DualPhase_1}(a). It was operated at 40.00 kV (mean energy of 28.00 keV) with continuous fluoroscopy mode with 0.40 mm nominal focal spot. The X-ray tube current was set at 12.5 mA, with a 5.00 second exposure period for each phase step. The X-ray detector is a direct conversion type a-Se detector (AXS-2430, Analogic Corporation, Quebec, Canada) with a native element dimension of 85.00 $\mu$m. The detector is tilted by 80 degree to generate an effective detector pixel size of 14.70 $\mu m$. The source grating $G_0$ has a period of 24.00 $\mu m$, with a duty cycle of 0.35. The first phase grating $G_1$ has a period of 4.364 $\mu m$, with a duty cycle of 0.50, and the second phase grating $G_2$ has a period of 4.640 $\mu m$, with a duty cycle of 0.50. With these available gratings, the best imaging distance between $G_0$ and $G_1$ is 527.86 mm, 108.90 mm between $G_1$ and $G_2$, and 1627.74 mm between $G_2$ and the detection plane. Moreover, the distance between the X-ray tube focal spot and the $G_0$ is 40.00 mm. The total phase stepping number of the $G_0$ grating is set to six, with a stepping interval of 4.00 $\mu m$.

With the same grating setup, another validation experiment was also performed by interchanging the X-ray focal spot position and the detector position, see Fig.~\ref{fig:Fig_DualPhase_1}(b). In particular, the Varex G242 tube was replaced by an X-ray micro-focus tube (UltraBright 96000, Oxford Instruments, CA, USA) with $80.00 \mu m$ focal spot size.

\section{Results}
\subsection{Theoretical results}
\subsubsection{Intensity of diffraction pattern}
As a result, according to the Eq.~(\ref{eq:ch01_Fresnel_Kirchhoff_4}), the X-ray field intensity at the detector plane is expressed as:
\begin{align}
\label{eq:ch01_Fresnel_Kirchhoff_5}
I_{3}(x_{3})=&\frac{U_{0}^{2}}{(d_{1}+d_{2}+d_{3})^{2}}\sum_{s=-\infty}^{s=\infty}\sum_{r=-\infty}^{r=\infty}C_{s}\left(\frac{d_{1}(d_{2}+d_{3})s+d_{1}d_{3}\frac{p_{1}}{p_{_{2}}}r}{d_{1}+d_{2}+d_{3}},\lambda,p_{1},\phi_{1}\right) \\ \nonumber
&C_{r}\left(\frac{(d_{1}+d_{2})d_{3}r+d_{1}d_{3}\frac{p_{2}}{p_{1}}s}{d_{1}+d_{2}+d_{3}},\lambda,p_{2},\phi_{2}\right)sinc\left(\frac{[d_{3}\frac{r}{p_{2}}+(d_{2}+d_{3})\frac{s}{p_{1}}]\sigma}{d_{1}+d_{2}+d_{3}}\right) \\  \nonumber
&e^{\frac{i2\pi[d_{3}\frac{r}{p_{2}}+(d_{2}+d_{3})\frac{s}{p_{1}}]x_{s}}{d_{1}+d_{2}+d_{3}}}
sinc\left(\frac{[(d_{1}+d_{2})\frac{r}{p_{2}}+d_{1}\frac{s}{p_{1}}]p_{del}}{d_{1}+d_{2}+d_{3}}\right)e^{\frac{i2\pi[(d_{1}+d_{2})\frac{r}{p_{2}}+d_{1}\frac{s}{p_{1}}]x_{3}}{d_{1}+d_{2}+d_{3}}},
\end{align}
where $p_1$ and $p_2$ corresponds to the period of the first and second phase grating, $s,r$ denotes their diffraction order correspondingly, $\phi_{1}$ and $\phi_{2}$ are the X-ray wave-front phase shifts induced on the two phase gratings, $p_{del}$ corresponds to the dimension of detector element, $x_s$ denotes the $x$-coordinate of the source plane, $x_3$ denotes the $x$-coordinate of the detector plane. In this paper, we focus on discussing the two $\pi-\pi$ phase gratings based X-ray interferometer, i.e., $\phi_{1}=\pi$ and $\phi_{2}=\pi$. Hence\cite{yan2015a, yan2016predicting},
\begin{align}
& C_{s}\left(\frac{d_{1}(d_{2}+d_{3})s+d_{1}d_{3}\frac{p_{1}}{p_{_{2}}}r}{d_{1}+d_{2}+d_{3}},\lambda,p_{1},\pi\right)=-\frac{2}{\pi}sin\left(\frac{2\pi[d_{1}(d_{2}+d_{3})\frac{s}{p_{1}}+d_{1}d_{3}\frac{r}{p_{_{2}}}]\lambda}{(d_{1}+d_{2}+d_{3})p_{1}}\right),
\label{eq:ch01_Fresnel_Kirchhoff_6}
\end{align}
and
\begin{align}
& C_{r}\left(\frac{(d_{1}+d_{2})d_{3}r+d_{1}d_{3}\frac{p_{2}}{p_{1}}s}{d_{1}+d_{2}+d_{3}},\lambda,p_{2},\pi\right)=-\frac{2}{\pi}sin\left(\frac{2\pi[(d_{1}+d_{2})d_{3}\frac{r}{p_{2}}+d_{1}d_{3}\frac{s}{p_{2}}]\lambda}{(d_{1}+d_{2}+d_{3})p_{2}}\right).
\label{eq:ch01_Fresnel_Kirchhoff_7}
\end{align}
Substituting Eq.~(\ref{eq:ch01_Fresnel_Kirchhoff_6}) and Eq.~(\ref{eq:ch01_Fresnel_Kirchhoff_7}) back into Eq.~(\ref{eq:ch01_Fresnel_Kirchhoff_5}), and further assuming only the lowest order, i.e., $s=-2$ and $r=2$, or $s=2$ and $r=-2$, dominates the contribution to the detectable diffraction fringe, the following equation can be derived:
\begin{align}
 \label{eq:ch01_Fresnel_Kirchhoff_8}
&I_{3}(x_{3})=\frac{U_{0}^{2}}{(d_{1}+d_{2}+d_{3})^{2}}\bigg(1+\frac{8}{\pi^{2}}sin\left(\frac{4\pi d_{1}[(d_{2}+d_{3})-d_{3}\frac{p_{1}}{p_{_{2}}}]}{(d_{1}+d_{2}+d_{3})Z_{t,1}}\right)sin\left(\frac{4\pi d_{3}[(d_{1}+d_{2})-d_{1}\frac{p_{2}}{p_{1}}]}{(d_{1}+d_{2}+d_{3})Z_{t,2}}\right) \\ \nonumber
 &sinc\left(\frac{2[d_{2}+d_{3}-d_{3}\frac{p_{1}}{p_{2}}]\sigma}{(d_{1}+d_{2}+d_{3})p_{1}}\right)
 e^{\frac{i4\pi[(d_{2}+d_{3})-d_{3}\frac{p_{1}}{p_{_{2}}}]x_s}{(d_{1}+d_{2}+d_{3})p_1}}
 sinc\left(\frac{2[(d_{1}+d_{2})-d_{1}\frac{p_{2}}{p_{1}}]p_{del}}{(d_{1}+d_{2}+d_{3})p_{2}}\right)e^{\frac{i4\pi[(d_{1}+d_{2})-d_{1}\frac{p_{2}}{p_{1}}]x_{3}}{(d_{1}+d_{2}+d_{3})p_{2}}}\bigg),
\end{align}
where $Z_{t,1}=\frac{p_1^2}{\lambda},Z_{t,2}=\frac{p_2^2}{\lambda}$.

\begin{figure*}[t]
\centering
\includegraphics[width=0.85\textwidth]{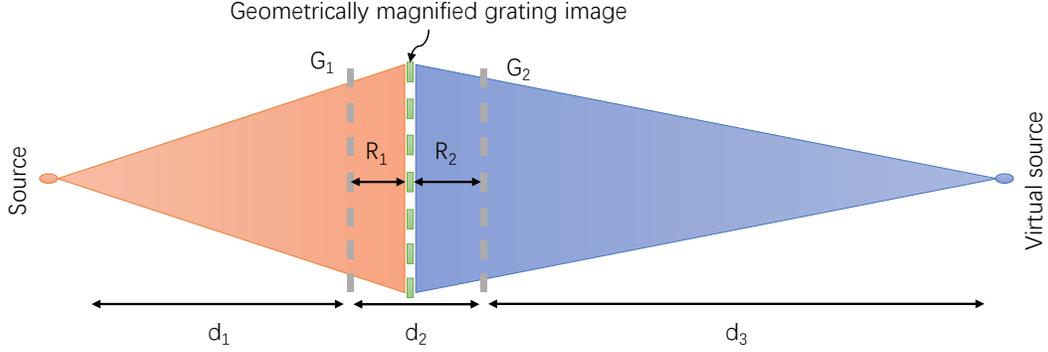}
\vspace{-0.05in}
\caption{Illustration of the virtual image plane, which ensures that the periods of the geometrically magnified image of $G_{1}$ (when x-ray emitted from the source on the left) and $G_{2}$ (when x-ray emitted from the virtual source on the right) are identical.}
\label{fig:fig2} 
\end{figure*}

\subsubsection{Geometrical interpretation}
In one of the previous studies\cite{kagias2017dual,lei2018cascade}, Kagias et. al. made an assumption that a virtual image plane exists between the first and second phase grating. Such idea was also utilized in another experiment conducted by Lei et. al. Unfortunately, so far this assumption has not been demonstrated in theory. Following the virtual image plane idea, we try to provide a more intuitive interpretation of the complex mathematical expression in Eq.~(\ref{eq:ch01_Fresnel_Kirchhoff_8}). As shown in Fig.~\ref{fig:fig2}, supposing there is a virtual plane lies between $G_{1}$ and $G_{2}$ such that it ensures the periods of the geometrically magnified image of $G_{1}$ (when x-ray emitted from the source) and $G_{2}$ (when x-ray emitted from the virtual source) are identical, namely,
\begin{align}
\frac{d_{1}+R_{1}}{d_{1}}p_{1}=\frac{d_{3}+R_{2}}{d_{3}}p_{2}.
 \label{eq:ch01_Fresnel_Kirchhoff_9}
\end{align}
Essentially, the above assumption is made based on the principle of optical reversibility, which presupposes that the attenuation of a light ray during its passage through an optical medium because of reflection, refraction, and absorption is not affected by a reversal of the direction of the ray. From Eq.~(\ref{eq:ch01_Fresnel_Kirchhoff_9}), both $R_1$ and $R_2$ can be obtained as below:
\begin{align}
R_{1}&=\frac{d_{1}(d_{2}+d_{3}-d_{3}\frac{p_{1}}{p_{2}})}{d_{3}\frac{p_{1}}{p_{2}}+d_{1}}, \\
R_{2}&=\frac{d_{2}(d_{1}+d_{2}-d_{1}\frac{p_{2}}{p_{1}})}{d_{3}+d_{1}\frac{p_{2}}{p_{1}}}.
\label{eq:R1R2}
\end{align}
With some mathematical derivations, it is easy to get the following relationships:
\begin{align}
\label{eq:f1}
f_{1}:=\frac{d_{1}(d_{2}+d_{3}-d_{3}\frac{p_{1}}{p_{2}})}{(d_{1}+d_{2}+d_{3})}&=\frac{R_{1}d_{1}}{R_{1}+d_{1}}=\frac{1}{\frac{1}{d_1}+\frac{1}{R_1}}, \\
\label{eq:f2}
f_{2}:=\frac{d_{3}(d_{1}+d_{2}-d_{1}\frac{p_{2}}{p_{1}})}{(d_{1}+d_{2}+d_{3})}&=\frac{R_{2}d_{3}}{R_{2}+d_{3}}=\frac{1}{\frac{1}{d_3}+\frac{1}{R_2}},
\end{align}
which could be considered an analogue to the thin lens equation. Herein, $f_1$ and $f_2$ are the focal length of the $G_1$ and $G_2$ grating, correspondingly. If the X-ray beams propagate from left to right, see Fig.~\ref{fig:fig2}, then $d_1$ is the object distance, and $R_1$ is the image distance. Regarding to the $G_2$ grating, whereas, $R_2$ is the object distance, and $d_3$ is the image distance. Therefore, Eq.~(\ref{eq:ch01_Fresnel_Kirchhoff_8}) can be simplified into
\begin{align}
 I_{3}(x_{3})&=\frac{U_{0}^{2}}{(d_{1}+d_{2}+d_{3})^{2}}\bigg(1+\frac{8}{\pi^{2}}sin\left(\frac{4\pi f_{1}}{Z_{t,1}}\right)sin\left(\frac{4\pi f_{2}}{Z_{t,2}}\right)sinc\left(\frac{2f_{1}\sigma}{d_{1}p_{1}}\right)e^{\frac{i4\pi f_{1}x_{s}}{d_{1}p_{1}}}sinc\left(\frac{2f_{2}p_{del}}{d_{3}p_{2}}\right)e^{\frac{i4\pi f_{2}x_{3}}{d_{3}p_{2}}}\bigg).
\label{eq:ch01_Fresnel_Kirchhoff_10}
\end{align}

\subsubsection{Fringe visibility}
From Eq.~(\ref{eq:ch01_Fresnel_Kirchhoff_10}), the derived fringe visibility, denoted as $\epsilon$, is found to be:
\begin{align}
\epsilon = \frac{8}{\pi^{2}}sin\left(\frac{4\pi f_{1}}{Z_{t,1}}\right)sin\left(\frac{4\pi f_{2}}{Z_{t,2}}\right)sinc\left(\frac{2f_{1}\sigma}{d_{1}p_{1}}\right)sinc\left(\frac{2f_{2}p_{del}}{d_{3}p_{2}}\right).
\label{eq:ch01_Fresnel_Kirchhoff_11}
\end{align}
 In Eq.~(\ref{eq:ch01_Fresnel_Kirchhoff_11}), the first two sine functions are oscillated terms, the third and fourth terms can be considered as two slowly decayed terms (this is true especially for their main lobes). Therefore, locally optimal visibility can be achieved whenever the two sine functions reach their maximum or minimum values, namely,
\begin{align}
f^{op}_1 & = \frac{2l_1+1}{8}Z_{t,1}=\frac{2l_1+1}{8}\frac{p_1^2}{\lambda},\;\;\; l_1\in\mathbb{Z}\\
f^{op}_2 & = \frac{2l_2+1}{8}Z_{t,2}=\frac{2l_2+1}{8}\frac{p_2^2}{\lambda},\;\;\; l_2\in\mathbb{Z}
\end{align}
Notice that the source size $\sigma$ and the detector element size $p_{del}$ are presumed to be fixed during these analyses. Interestingly, these optimal focal length $f^{op}_1$ and $f^{op}_2$ are exactly the self-image distance of a standard Talbot interferometer. If $l_1$ and $l_2$ are equal to 0, the entire system setup becomes the most compact. Based on Eq.~(\ref{eq:f1}) and Eq.~(\ref{eq:f2}), additionally, the optimal values for $R^{op}_1$ and $R^{op}_2$ can be determined immediately:
\begin{align}
\label{eq:R1}
R^{op}_1 & = \frac{f^{op}_1 d_1}{d_1-f^{op}_1},\\
\label{eq:R2}
R^{op}_2 & = \frac{f^{op}_2 d_3}{d_3-f^{op}_2}.
\end{align}

\subsubsection{Periods of source grating and fringe}
Based on Eq.~(\ref{eq:ch01_Fresnel_Kirchhoff_10}), the pattern period on the source plane is found to be:
\begin{align}
p_{0}&=\frac{d_{1}p_{1}}{2f_{1}}=\frac{d_{1}+d_{2}+d_{3}}{2(\frac{d_{2}+d_{3}}{p_{1}}-\frac{d_{3}}{p_{2}})},
\end{align}
where $p_0$ corresponds to the source grating period when a medical grade diagnostic X-ray tube is utilized. In addition, the fringe period $p_f$ on the detector plane is equal to:
\begin{align}
p_{f}&=\frac{d_{3}p_{2}}{2f_{2}}=\frac{d_{1}+d_{2}+d_{3}}{2(\frac{d_{1}+d_{2}}{p_{2}}-\frac{d_{1}}{p_{1}})}.
\end{align}
Combining the above two equations, one can readily get
\begin{align}
\label{eq:p0_2}
p_{0}=\frac{p_{f}p_{1}p_{2}}{2p_{f}(p_{2}-p_{1})+p_{1}p_{2}},\\
\label{eq:pf_2}
p_{f}=\frac{p_{0}p_{1}p_{2}}{2p_{0}(p_{1}-p_{2})+p_{1}p_{2}}.
\end{align}
Clearly, if the two phase gratings have identical periods, i.e., $p_1 = p_2$, the detected fringe period should be equal to the source grating period, i.e., $p_f = p_0$. To ease the fringe detection, in reality, one might want the diffraction fringe has a larger period than the source grating period, i.e., $p_f > p_0$, therefore, $p_1 < p_2$ should be satisfied. In other words, the $G_1$ grating should have a smaller period than the $G_2$ grating. Whenever such grating designs are used, in this paper we call it asymmetric dual-phase grating interferometer system. If the fringe period goes to infinite, then the three gratings should satisfy the following relationship,
\begin{align}
\frac{1}{p_0}-\frac{2}{p_1}+\frac{2}{p_2}=0.
\end{align}

\subsubsection{Key parameter predictions}
\begin{figure*}[t]
\centering
\includegraphics[width=1\textwidth]{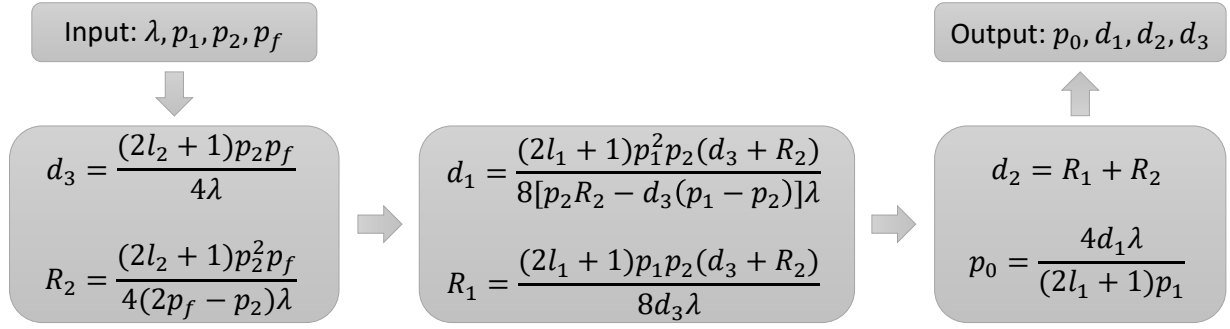}
\vspace{-0.05in}
\caption{Flowchart to estimate the key parameters of a dual-phase grating interferometer system when $\lambda$, $p_1$, $p_2$, and $p_f$ are known in prior.}
\label{fig:Fig_DualPhase_2}
\end{figure*}

By far, we have established a theoretical framework to predict all the key parameters needed by a dual-phase grating interferometer system. The Fig.~\ref{fig:Fig_DualPhase_2} summarized one example to perform such parameter estimations. Herein, we assume the already known parameters are the wavelength of the mean energy X-ray beam, the periods of the two phase gratings, and the expected period of the diffraction fringe. Following these calculations, the inter-space length, i.e., $d_1$, $d_2$, $d_3$, and the source grating period $p_0$ can all be obtained readily. If other parameters are known in prior, in principle, the rest parameters can also be predicted using our theory.

\subsection{Numerical simulation results}
\begin{figure*}[ht]
\centering
\includegraphics[width=1\textwidth]{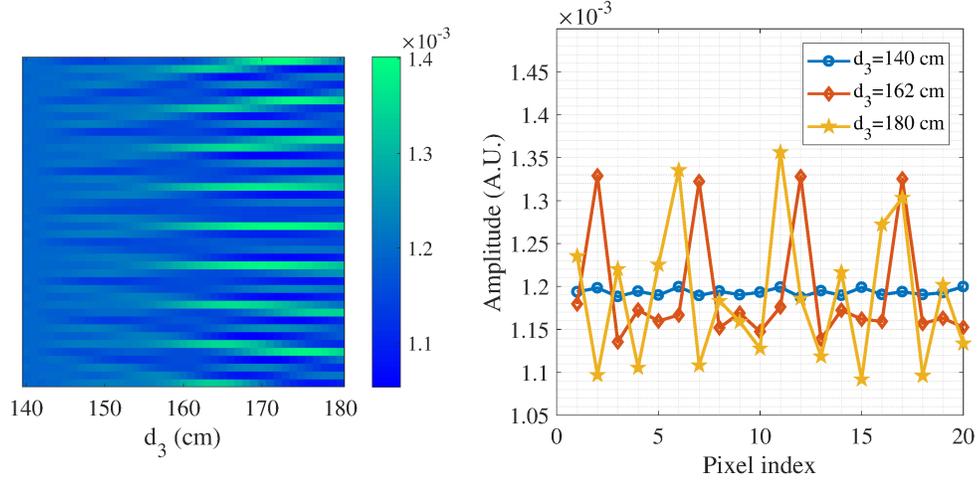}
\vspace{-0.05in}
\caption{Numerical simulation results. Image on the left shows the diffraction fringe distribution for $d_3$ between 140.00 mm to 180.00 mm. Plots on the right show the fringe line profile at three different $d_3$ positions.}
\label{fig:fig3} 
\end{figure*}

Numerical simulation results are shown in Fig.~\ref{fig:fig3}. As can be seen, the formed diffraction fringes become pronounced when $d_3 > 155 cm$. The line profiles also demonstrated this observation. For this simulation study, the fringe period at $d_3=162$ cm observation plane is close to 5 detector elements, corresponding to about $70.00 \mu m$ fringe period. This numerical result agrees with the theoretical prediction. At this particular location, the numerically obtained fringe visibility is around $8\%$. The relatively lowered fringe visibility is mainly due to the used polychromatic X-ray beam.

\subsection{Experimental results}
\begin{figure*}[t]
\centering
\includegraphics[width=1\textwidth]{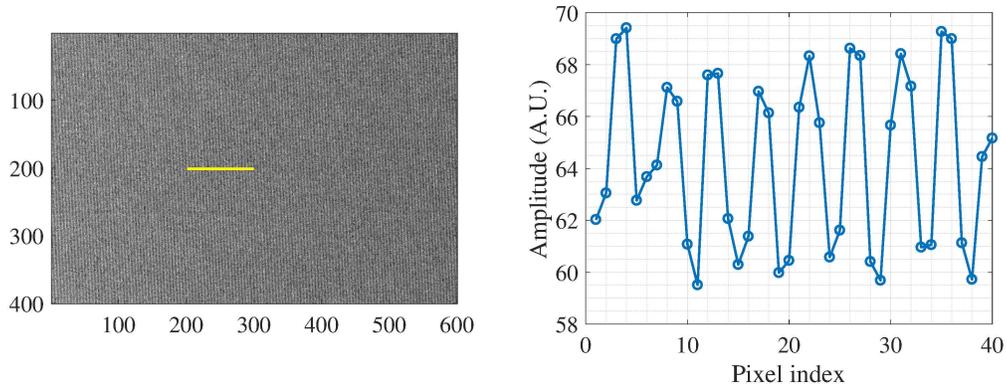}
\vspace{-0.15in}
\caption{Experimental results of system setup in Fig.~2(a). Image on the left corresponds to the detected fringes, and the plot on the right hand side represents the line profile of the fringe. In this case, the effective pixel dimension along the horizontal direction is $14.70 \mu m$.}
\label{fig:exp1} 
\end{figure*}

\begin{figure*}[t]
\centering
\includegraphics[width=1\textwidth]{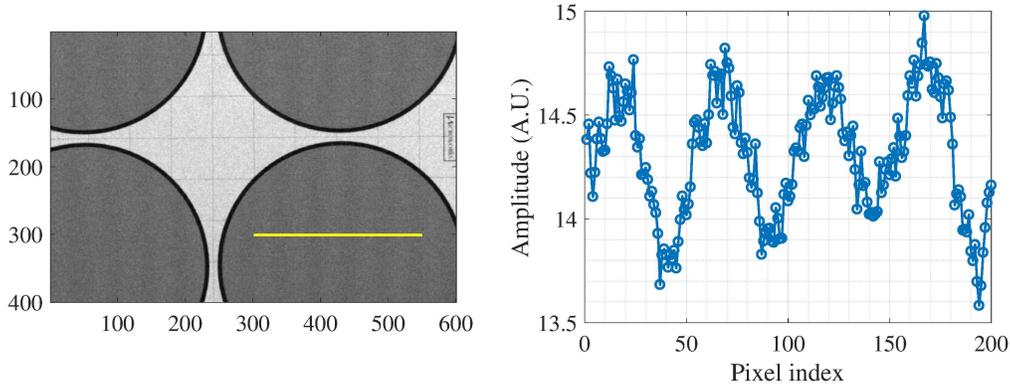}
\vspace{-0.15in}
\caption{Experimental results of system setup in Fig.~2(b). Image on the left corresponds to the detected fringes, and the plot on the right hand side represents the line profile of the fringe. In this case, the effective pixel dimension along the horizontal direction is $85.00 \mu m$.}
\label{fig:exp2} 
\end{figure*}

Experimental results of the illustrated two system setups in Fig.~\ref{fig:Fig_DualPhase_1}(a)-(b) are presented in Fig.~\ref{fig:exp1} and Fig.~\ref{fig:exp2}, correspondingly. For system setup in Fig.~\ref{fig:Fig_DualPhase_1}(a), the detected fringe period is about $73.50 \mu m$, with fringe visibility around $8\%$. The experimental results agree well with the numerical simulation results shown in Fig.~\ref{fig:fig3}. For system setup in Fig.~\ref{fig:Fig_DualPhase_1}(b), the detected Moir\'{e} fringe period is about $4.25 mm$, with fringe visibility close to $3\%$. The drop of this detected fringe visibility is mainly due to the relatively large focal spot size ($80.00 \mu m$) of our aged microfocus tube.

\section{Discussion}
In order to realize clinical X-ray dual-phase grating differential phase contrast imaging, a source grating is usually needed to increase the beam coherence of a diagnostic grade X-ray source with large focal spot size. However, previous experimental work and theoretical work did not provide sufficient proofs in estimating the period of the source grating. To overcome this problem, we established a theoretical analyses framework based on X-ray wave optics. Not only being able to predict the source grating period, this theoretical framework is also capable of estimating other parameters needed by a dual-phase grating interferometer system, for example, the inter-grating distances, the diffraction fringe period, and the phase grating periods.

In principle, the two phase gratings can be designed to have any period if there is no source grating added. However, when a source grating is used, our theory indicates that asymmetric dual-phase grating interferometer should be a better choice in generating diffraction fringes with large period. Herein, asymmetric means that the two phase gratings have different grating periods. Specifically, the period of the $G_1$ grating should slightly less than the period of the $G_2$ grating, assuming that the X-ray beam passes through the $G_1$ grating first. Under such condition, one is able to manipulate the period of the formed diffraction fringe to be detectable by a certain detector. If the $G_1$ and $G_2$ gratings have identical periods, the formed diffraction fringe should always have identical period as of the source grating. This might not always good for practical applications because source grating with small period ($<30 \mu m$) ensures high tube output usage efficiency, however, such small fringe period ($<30 \mu m$) obviously does not ease its detection by normal flat panel detector. Therefore, asymmetric grating design helps to make a trade-off between the tube output usage efficiency and the fringe period.

In our theory, the period of the source grating and the period of the diffraction fringe could be treated as two counterparts. Without specifying the X-ray beam propagation direction, it is hard to distinguish them from the derived formula, see Eq.~(\ref{eq:ch01_Fresnel_Kirchhoff_5}) and Eq.~(\ref{eq:ch01_Fresnel_Kirchhoff_8}). Upon such observation, we believe this implies an inherent optical symmetry of a grating based X-ray interferometer. Thus, for the first time we are able to explain the dual-phase grating imaging theory via a geometrical approach, which is similar to the standard lens imaging theory. With this new scenario, the two phase gratings are considered as two thin convex lens. As a result, an image of the X-ray source (arrayed or single) is formed at the $R_1$ distance downstream of the $G_1$ grating, and it is imaged again by the $G_2$ grating to form the final image (diffraction fringe) on the detector plane. In general, the distance between $G_1$ and $G_2$ gratings can be arbitrary. However, certain conditions need to be satisfied to generate diffraction patterns with maximal fringe visibility. Our studies demonstrate that this happens when the defined effective optical lengths of $G_1$ and $G_2$ gratings are equal to their Talbot self-image distances. When the first ordered Talbot self-image distances are selected, the dual-phase grating interferometer system will have the most compact geometry. If the inter-distance between the $G_1$ and $G_2$ gratings does not satisfy the derived optimal conditions, see Eq.~(\ref{eq:R1}) and Eq.~(\ref{eq:R2}), both the $f_1$ and $f_2$ values should be adapted correspondingly according to Eq.~(\ref{eq:f1}) and Eq.~(\ref{eq:f2}). For this case, the ultimate fringe visibility will be degraded.

Ideally, the fringes generated from the two experiments shown in Fig.~\ref{fig:Fig_DualPhase_1} should have similar visibility, however, our results did not obtain the similar fringe visibility for the two experiments. Such mismatch is mainly caused by the large focal spot size ($80.00 \mu m$) of our aged Oxford microfocus tube, whose initial focal spot size is $20.00 \mu m$ and has been used for ten years.

This study has several limitations: first, we only focused on the $\pi-\pi$ dual-phase grating system, and results for the $\frac{\pi}{2}-\frac{\pi}{2}$ dual-phase grating system are not presented in this paper. However, results for the latter system settings should be able to be derived readily, and similar conclusions can also be obtained. Second, no quantitative results have obtained for the system sensitivity in current theoretical framework, and this would be investigated in future. As can be expected, the system design and optimization should take the interferometer sensitivity factor into consideration as well.

\section{Conclusion}
In conclusion, in this paper we provide a theoretical analysis framework based on wave optics to precisely predict the key parameters of a dual-phase grating interferometer, particularly the period of the source grating when a medical grade X-ray tube with large focal spot is used. Similar to the thin lens imaging, we also provide a geometrical explanation to the obtained dual-phase grating imaging theory. The high agreements between the numerical and experimental studies demonstrate the correctness of the theory.

\section*{Acknowledgments}
The authors would like to thank Dr. Zhicheng Li at Shenzhen Institutes of Advanced Technology, Chinese Academy of Sciences, for lending phase gratings.

\section*{Funding}
This project is supported by the National Natural Science Foundation of China (Grant No.~11804356, 11535015, 11674232), and Shenzhen Basic Research Program (JCYJ20170413162354654).

\section*{Appendix}
Using Eq.~(\ref{eq:ch01_trans_f_1}) and Eq.~(\ref{eq:ch01_Fresnel_Kirchhoff_4}), the X-ray wave field disturbance right before the $G_1$ grating is obtained as following:
\begin{align}
 & \text{U}_{1}(x_{1},y_{1})=\frac{\text{U}_{0}e^{ikd_{1}}}{d_{1}}e^{ik\left[\frac{(x_{1}-x_{s})^{2}}{2d_{1}}+\frac{(y_{1}-y_{s})^{2}}{2d_{1}}\right]},
\label{eq:ch01_onegrating_1}
\end{align}
where $(x_s, y_s)$ denotes the source coordinates. Afterwards, the X-ray field interacts with $G_1$ grating via
\begin{align}
 & \text{U}_{1}^{'}(x_{1},y_{1})=\text{U}_{1}(x_{1},y_{1})\text{T}_{1}(x_{1};p_1).
 \label{eq:ch01_onegrating_4}
 \end{align}
Then, the X-ray field reaches the $G_2$ grating with field disturbance of
\begin{align}
 &\text{U}_{2}(x_{2},y_{2})=\frac{U_{0}e^{ik(d_{1}+d_{2})}}{d_{1}+d_{2}}\sum_{n=-\infty}^{n=\infty}a_{n}e^{\frac{i2\pi n(-d_{1}n\pi+kp_{1}x_{s})}{kp_{1}^{2}}}e^{ik\left[\frac{[x_{2}-(x_{s}-\frac{2\pi d_{1}n}{kp_{1}})]^{2}}{2(d_{1}+d_{2})}+\frac{(y_{2}-y_{s})^{2}}{2(d_{1}+d_{2})}\right]}. 
 \label{eq:ch01_onegrating_6}
\end{align}
Similarly, X-ray field starts to interact with $G_2$ grating,
\begin{align}
 & \text{U}_{2}^{'}(x_{2},y_{2})=\text{U}_{2}(x_{2},y_{2})\text{T}_{2}(x_{2};p_2).
 \label{eq:ch01_twograting_4}
\end{align}
Finally, the X-ray wave arrives at the detector plane with field disturbance written as below:
\begin{align}
  \text{U}_{3}(x_{3},y_{3})=&\frac{U_{0}e^{ik(d_{1}+d_{2}+d_{3})}}{d_{1}+d_{2}+d_{3}}\sum_{n=-\infty}^{n=\infty}\sum_{m=-\infty}^{m=\infty}a_{n}b_{m}e^{\frac{i2\pi n(-d_{1}n\pi+kp_{1}x_{s})}{kp_{1}^{2}}}e^{\frac{i2\pi m[-(d_{1}+d_{2})m\pi+kp_{2}(x_{s}-\frac{2\pi d_{1}n}{kp_{1}})]}{kp_{2}^{2}}}\nonumber \\&e^{ik\left[\frac{[x_{3}-(x_{s}-\frac{2\pi d_{1}n}{kp_{1}}-\frac{2\pi(d_{1}+d_{2})m}{kp_{2}})]^{2}}{2(d_{1}+d_{2}+d_{3})}+\frac{(y_{3}-y_{s})^{2}}{2(d_{1}+d_{2}+d_{3})}\right]}.
\label{eq:ch01_twograting_6}
\end{align}
Therefore, the beam intensity can be expressed as:
\begin{align}
  I_{3}(x_{3},y_{3})=&\frac{U_{0}^{2}}{(d_{1}+d_{2}+d_{3})^2}\sum_{n=-\infty}^{n=\infty}\sum_{n'=-\infty}^{n'=\infty}\sum_{m=-\infty}^{m=\infty}\sum_{m'=-\infty}^{m'=\infty}a_{n}a_{n'}^{*}b_{m}b_{m'}^{*}\nonumber \\&e^{-\frac{i\pi[(d_{1}+d_{2})d_{3}(m^{2}-m'^{2})p_{1}^{2}+2d_{1}d_{3}(mn-m'n')p_{1}p_{2}+d_{1}(d_{2}+d_{3})(n^{2}-n'^{2})p_{2}{}^{2}]\lambda}{(d_{1}+d_{2}+d_{3})p_{1}^{2}p_{2}{}^{2}}}\nonumber \\&e^{\frac{i2\pi[d_{3}(m-m')p_{1}+(d_{2}+d_{3})(n-n')p_{2}]x_{s}}{(d_{1}+d_{2}+d_{3})p_{1}p_{2}}}e^{\frac{i2\pi[(d_{1}+d_{2})(m-m')p_{1}+d_{1}(n-n\text{'})p_{2}]x_{3}}{(d_{1}+d_{2}+d_{3})p_{1}p_{2}}}.
 \label{eq:ch01_onegrating_7}
\end{align}
Let $n=n'+s,m=m'+r$, Eq.~(\ref{eq:ch01_onegrating_7}) can be simplified into
\begin{align}
  I_{3}(x_{3},y_{3})=&\frac{U_{0}^{2}}{(d_{1}+d_{2}+d_{3})^{2}}\sum_{n'+s=-\infty}^{n'+s=\infty}\sum_{n'=-\infty}^{n'=\infty}\sum_{m'+r=-\infty}^{m'+r=\infty}\sum_{m'=-\infty}^{m'=\infty}a_{n'+s}a_{n'}^{*}b_{m'+r}b_{m'}^{*}\nonumber \\&e^{-\frac{i2\pi s(2n+s)\lambda}{\frac{2(d_{1}+d_{2}+d_{3})}{d_{1}(d_{2}+d_{3})+d_{1}d_{3}\frac{p_{1}}{p_{2}}\frac{r}{s}}p_{1}^{2}}}e^{-\frac{i2\pi r(2m+r)\lambda}{\frac{2(d_{1}+d_{2}+d_{3})}{(d_{1}+d_{2})d_{3}+d_{1}d_{3}\frac{p_{2}}{p_{1}}\frac{s}{r}}p_{2}^{2}}}\nonumber \\&e^{-\frac{i2\pi[(d_{2}+d_{3})\frac{s}{p_{1}}+d_{3}\frac{r}{p_{2}}]x_{s}}{(d_{1}+d_{2}+d_{3})}}e^{-\frac{i2\pi[(d_{1}+d_{2})\frac{r}{p_{2}}+d_{1}\frac{s}{p_{1}}]x_{3}}{(d_{1}+d_{2}+d_{3})}}.
  \label{eq:ch01_onegrating_8}
\end{align}
Using equation\cite{arrizon1992irradiance, yan2015a, yan2016predicting}
\begin{align}
\sum_{l=-\infty}^{l=\infty} C_{l}(d,\lambda,p_{g},\Delta\phi_{g})=\sum_{l=-\infty}^{l=\infty}\sum_{n=-\infty}^{n=\infty}g_{l+n}g_{n}^{*}e^{\frac{-i2\pi l(l+2n)\lambda d}{2p_{g}^{2}}},
\end{align}
the Eq.~(\ref{eq:ch01_onegrating_8}) can be further simplified as following,
\begin{align}
I_{3}(x_{3},y_{3})=&\frac{U_{0}^{2}}{(d_{1}+d_{2}+d_{3})^{2}}\sum_{s=-\infty}^{s=\infty}\sum_{r=-\infty}^{r=\infty}C_{s}(\frac{d_{1}(d_{2}+d_{3})+d_{1}d_{3}\frac{p_{1}}{p_{_{2}}}\frac{r}{s}}{d_{1}+d_{2}+d_{3}},\lambda,p_{1},\phi_{1})
\\ \nonumber
&C_{r}(\frac{(d_{1}+d_{2})d_{3}+d_{1}d_{3}\frac{p_{2}}{p_{1}}\frac{s}{r}}{d_{1}+d_{2}+d_{3}},\lambda,p_{2},\phi_{2})
e^{\frac{i2\pi[d_{3}\frac{r}{p_{2}}+(d_{2}+d_{3})\frac{s}{p_{1}}]x_{s}}{d_{1}+d_{2}+d_{3}}}
e^{\frac{i2\pi[(d_{1}+d_{2})\frac{r}{p_{2}}+d_{1}\frac{s}{p_{1}}]x_{3}}{d_{1}+d_{2}+d_{3}}}.
\label{eq:ch01_onegrating_9}
\end{align}
where $\phi_{1}$ and $\phi_{2}$ denotes the phase shift on $G_1$ and $G_2$ grating correspondingly.

Now, we take the source size and the detector pixel size into consideration. Assuming the source has a slit shape and is defined by Eq.~(\ref{eq:source_1}), the detector element has a dimension of $p_{del}$, thus, Eq.~(\ref{eq:ch01_onegrating_9}) becomes
\begin{align}
I_{3}(x_{3},y_{3})=&\frac{U_{0}^{2}}{(d_{1}+d_{2}+d_{3})^{2}}\sum_{s=-\infty}^{s=\infty}\sum_{r=-\infty}^{r=\infty}C_{s}(\frac{d_{1}(d_{2}+d_{3})s+d_{1}d_{3}\frac{p_{1}}{p_{_{2}}}r}{d_{1}+d_{2}+d_{3}},\lambda,p_{1},\phi_{1})
\\ \nonumber
&C_{r}(\frac{(d_{1}+d_{2})d_{3}r+d_{1}d_{3}\frac{p_{2}}{p_{1}}s}{d_{1}+d_{2}+d_{3}},\lambda,p_{2},\phi_{2})
 \frac{1}{\sigma}\int_{x_{s}-\frac{\sigma}{2}}^{x_{s}+\frac{\sigma}{2}}e^{-\frac{i2\pi[(d_{2}+d_{3})\frac{s}{p_{1}}+d_{3}\frac{r}{p_{2}}]t}{(d_{1}+d_{2}+d_{3})}}dt\\ \nonumber
 & \frac{1}{P_d}\int_{x_{3}-\frac{P_{d}}{2}}^{x_{3}+\frac{P_{d}}{2}}e^{-\frac{i2\pi[(d_{1}+d_{2})\frac{r}{p_{2}}+d_{1}\frac{s}{p_{1}}]v}{(d_{1}+d_{2}+d_{3})}}dv,
\label{eq:ch01_onegrating_10}
\end{align}
namely,
\begin{align}
I_{3}(x_{3},y_{3})=&\frac{U_{0}^{2}}{(d_{1}+d_{2}+d_{3})^{2}}\sum_{s=-\infty}^{s=\infty}\sum_{r=-\infty}^{r=\infty}C_{s}(\frac{d_{1}(d_{2}+d_{3})+d_{1}d_{3}\frac{p_{1}}{p_{_{2}}}\frac{r}{s}}{d_{1}+d_{2}+d_{3}},\lambda,p_{1},\phi_{1})
\\ \nonumber
&C_{r}(\frac{(d_{1}+d_{2})d_{3}+d_{1}d_{3}\frac{p_{2}}{p_{1}}\frac{s}{r}}{d_{1}+d_{2}+d_{3}},\lambda,p_{2},\phi_{2})sinc(\frac{[d_{3}\frac{r}{p_{2}}+(d_{2}+d_{3})\frac{s}{p_{1}}]\sigma}{d_{1}+d_{2}+d_{3}})
\\ \nonumber
&e^{\frac{i2\pi[d_{3}\frac{r}{p_{2}}+(d_{2}+d_{3})\frac{s}{p_{1}}]x_{s}}{d_{1}+d_{2}+d_{3}}}
sinc(\frac{[(d_{1}+d_{2})\frac{r}{p_{2}}+d_{1}\frac{s}{p_{1}}]P_{d}}{d_{1}+d_{2}+d_{3}})e^{\frac{i2\pi[(d_{1}+d_{2})\frac{r}{p_{2}}+d_{1}\frac{s}{p_{1}}]x_{3}}{d_{1}+d_{2}+d_{3}}}.
\label{eq:ch01_onegrating_11}
\end{align}
Usually,
\begin{align*}
 & C_{l}(d,\lambda,p_{g},\Delta\phi_{g})=\left\{ \begin{array}{cc}
1 & l=0\\
-(1-cos\Delta\phi_{g})\cdot(-1)^{\lfloor4k\lambda d/p_{g}^{2}\rfloor}\cdot\frac{sin(4k^{2}\pi\lambda d/p_{g}^{2})}{k\pi} & l=2k\neq0\\
-i2sin\Delta\phi_{g}\cdot\frac{sin(4\pi\lambda d/p_{g}^{2}\cdot(k+1/2)^{2})}{\pi(2k+1)} & l=2k+1
\end{array}\right\} 
\end{align*}

%%%%%%%%%%%%%%%%%%%%%%% References %%%%%%%%%%%%%%%%%%%%%%%%%

%%%%%%%%%% If using BibTeX:
\bibliography{Bibliography_Paper}

%%%%%%%%%% If preparing manually:
% \begin{thebibliography}{1}
% \newcommand{\enquote}[1]{``#1''}

% \bibitem{Zhang:14}
% Y.~Zhang, S.~Qiao, L.~Sun, Q.~W. Shi, W.~Huang, L.~Li, and Z.~Yang,
%   \enquote{Photoinduced active terahertz metamaterials with nanostructured
%   vanadium dioxide film deposited by sol-gel method,}
%   {\protect\JournalTitle{Optics Express}} \textbf{22}, 11070--11078 (2014).

% \bibitem{OSA}
% {Optical Society}, \enquote{{OSA Publishing},}
%   \url{http://www.osapublishing.org}.

% \bibitem{FORSTER2007}
% P.~Forster, V.~Ramaswamy, P.~Artaxo, T.~Bernsten, R.~Betts, D.~Fahey,
%   J.~Haywood, J.~Lean, D.~Lowe, G.~Myhre, J.~Nganga, R.~Prinn, G.~Raga,
%   M.~Schulz, and R.~V. Dorland, \enquote{Changes in atmospheric consituents and
%   in radiative forcing,} in \enquote{Climate Change 2007: The Physical Science
%   Basis. Contribution of Working Group 1 to the Fourth assesment report of
%   Intergovernmental Panel on Climate Change,}  S.~Solomon, D.~Qin, M.~Manning,
%   Z.~Chen, M.~Marquis, K.~B. Averyt, M.~Tignor, and H.~L. Miler, eds.
%   (Cambridge University Press, 2007).

% \end{thebibliography}

\end{document}